\documentclass[reprint,aip,pop,amsmath,amssymb,showpacs]{revtex4-1}
\usepackage{graphicx}

\renewcommand{\a}{\alpha}
\renewcommand{\b}{\beta}
\newcommand{\de}{\delta}

\newcommand{\ve}{\varepsilon}

\renewcommand{\k}{\kappa}

\newcommand{\m}{\mu}
\newcommand{\n}{\nu}
\newcommand{\p}{\phi}
\newcommand{\vp}{\varphi}

\renewcommand{\r}{\rho}
\newcommand{\s}{\sigma}

\renewcommand{\Re}{\operatorname{Re}}

\renewcommand{\v}[1]{\mathbf{#1}}				
\newcommand{\unit}[1]{{\v{\hat{#1}}}}			

\newcommand{\eqnref}[1]{Eq.~\eqref{eqn:#1}}					

\newcommand{\comments}[1]{}									

\providecommand{\ol}{}		
\renewcommand{\ol}[1]{\overline{#1}}

\newcommand{\nablabarsq}{\ol{\nabla}^2}
\newcommand{\xbar}{{\ol{x}}}

\newcommand{\sinc}{\operatorname{sinc}}

\newcommand{\zf}{ZF}

\newcommand{\zfs}{ZFs}

\begin{document}

\title{Zonal Flow as Pattern Formation}

\author{Jeffrey B.~Parker}
\email{jbparker@princeton.edu}
\author{John A.~Krommes}
\email{krommes@princeton.edu}
\affiliation{Princeton Plasma Physics Laboratory, Princeton University, Princeton, New Jersey 08543, USA}

\date{\today}

\begin{abstract}
Zonal flows are well known to arise spontaneously out of turbulence. We show that for statistically averaged equations of the stochastically forced generalized Hasegawa-Mima model, steady-state zonal flows and inhomogeneous turbulence fit into the framework of pattern formation. There are many implications. First, the wavelength of the zonal flows is not unique. Indeed, in an idealized, infinite system, any wavelength within a certain continuous band corresponds to a solution. Second, of these wavelengths, only those within a smaller subband are linearly stable. Unstable wavelengths must evolve to reach a stable wavelength; this process manifests as merging jets.
\end{abstract}

\maketitle

Zonal flows (ZFs) --- azimuthally symmetric, generally banded, shear flows --- are spontaneously generated from turbulence and have been reported in atmospheric \cite{vasavada:2005} and laboratory plasma \cite{fujisawa:2009} contexts.  Recently, they have also been observed in astrophysical simulations.\cite{johansen:2009}  In magnetically confined plasmas, \zfs{} are thought to play a crucial role in regulation of turbulence and turbulent transport.\cite{lin:1998,diamond:2005}  A greater understanding of \zf{} behavior is valuable for untangling a host of nonlinear processes in plasmas, including details of transitions between modes of low and high confinement.

Zonal flows remain incompletely understood, even regarding the basic question of the jet width (wavelength).  In the plasma literature, one finds modulational or secondary instability calculations of \zf{} generation,\cite{diamond:2005,rogers:2000} but these cannot provide information on a saturated state.  Other theories typically make an assumption of long-wavelength \zfs{} and leave the \zf{} scale as an undetermined parameter.\cite{connaughton:2011}  Within geophysical contexts, various authors have attempted to relate the jet width or spacing to length scales that emerge from the vorticity equation by heuristically balancing the magnitudes of the Rossby wave term and the nonlinear advection.  Those scales include the Rhines scale and other, similar scales.\cite{rhines:1975,vallis:1993,sukoriansky:2007}  A Rhines-like length scale is also obtained from arguments based on potential vorticity staircases.\cite{dritschel:2008,scott:2012}  However, neither the heuristic Rhines estimates nor the paradigm of potential vorticity inversion and mixing generalize to more complex situations involving realistic plasma models.  We are therefore motivated to seek a more systematic approach to determining the \zf{} width that may offer such a generalization.

A related topic is the merging of jets.  Coalescence of two or more jets is ubiquitous in numerical simulations. \cite{huang:1998,scott:2007}  The merging process occurs during the initial transient period before a statistically steady state is reached.  It is clear that the merging is part of a dynamical process through which the \zf{} reaches its preferred length scale, but the merging phenomeon has not been understood thus far.

Our present work addresses these questions in the context of the stochastically forced generalized Hasegawa-Mima (GHM) equation, \cite{smolyakov:2000,krommes:2000} a model of magnetized plasma turbulence in the presence of a background density gradient.  This model is mathematically similar to the barotropic vorticity equation on a $\b$ plane. \cite{vallis:1993}  Our analysis is related to several recent works that focused on that equation in the geophysical context. \cite{srinivasan:2012,farrell:2003,farrell:2007,bakas:2013,constantinou:2012,marston:2008,tobias:2011,tobias:2013}  Importantly, numerical simulations of both models can display emergence of steady \zfs{}.  The GHM equation and the parameterizations of forcing and dissipation that we use are not realistic descriptions of plasma; however, the simplicity is an asset in understanding the qualitative behavior of these systems.

We study a statistical average of the flow.  Statistical approaches enable one to gain physical insight by averaging away the details of the turbulent fluctuations and working with smoothly varying quantities.  Sometimes statistical turbulence theories strive for quantitative accuracy, which requires rather complicated methods.\cite{krommes:2002}  In contrast, our investigation is at a more basic level and concerns the fundamental nature of \zfs{} interacting self-consistently with inhomogeneous turbulence.

Within the statistical framework, we build upon recent understanding of zonostrophic instability, in which homogeneous turbulence becomes unstable to \zf{} perturbations.\cite{srinivasan:2012}  Steady \zfs{} emerge from this bifurcation.  We show that the bifurcation obeys a classic amplitude equation, and therefore \zfs{} can be understood as pattern formation. \cite{cross:1993, cross:2009, hoyle:2006,clever:1974,newell:1990}  Two important results follow from the general properties of pattern-forming systems.  First, the wavelength of the \zf{} is not unique.  Indeed, in an idealized, infinite system, any wavelength within a certain continuous band corresponds to a steady-state solution.  Second, of these wavelengths, only those within a smaller subband are linearly stable.  Unstable wavelengths must evolve to reach a stable wavelength.  For unstable jets of short (long) wavelength, this process manifests as merging (branching) jets.

Our basic model is the 2D GHM equation,
	\begin{equation}
		\partial_t w(x,y) + \v{v} \cdot \nabla w - \k \partial_y \phi = \xi - \m w - \n (-1)^h \nabla^{2h} w,
		\label{eqn:GHMmodel}
	\end{equation}
where $\phi = (L_n / \r_s) e\vp / T_e$ is the normalized electrostatic potential, $L_n$ is the density gradient scale length, $\r_s$ is the sound radius, $T_e$ is the electron temperature, $w = \nabla^2 \phi - \hat{\a}\phi$ is the generalized vorticity and is related to ion gyrocenter density fluctuations $\de n_i^G$ by $w = -(L_n / \r_s) \de n_i^G / n_0$ where $n_0$ is the background density, $\hat{\a}$ is an operator such that in Fourier space $\hat{\a}(\v{k})=0$ if $k_y=0$ (\zf{} mode) and $\hat{\a}(\v{k}) = 1$ if $k_y\neq0$ (drift wave mode), $\v{v} = \unit{z} \times \nabla \phi$ is the $\v{E}\times\v{B}$ velocity, $\m$ is a constant frictional drag, $\n$ is the viscosity with hyperviscosity factor $h$, $\xi$ is white-noise forcing, and $\k$ is related to the density scale length.  Lengths are normalized to $\r_s$ and times are normalized to the drift wave period $\omega_*^{-1}=(L_n / \r_s) \Omega_i^{-1}$.  These normalizations and scalings are convenient to make $w$, $\phi$, and the active length and time scales of order unity, and they allow us to set $\k=1$.

\begin{figure}
		\centering
		\includegraphics{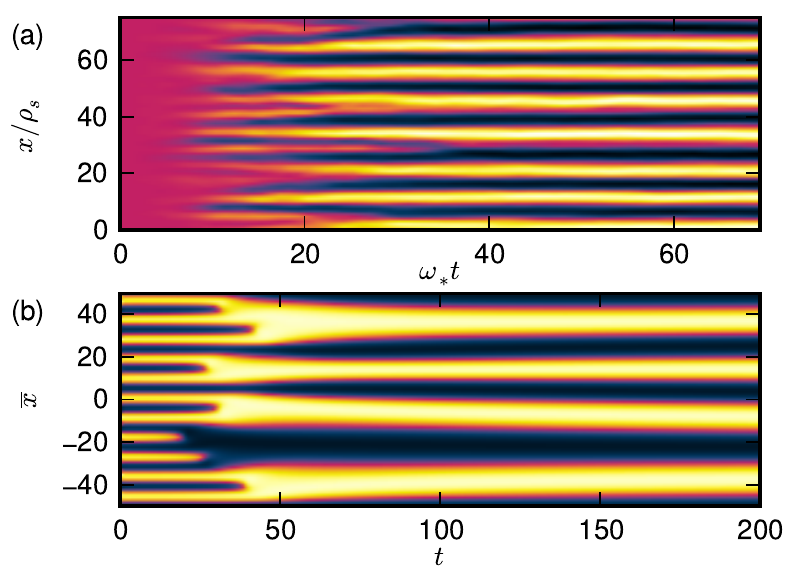}
		\caption{(a) Merging jets during the transient regime of equation \eqref{eqn:GHMmodel} (zonal-mean velocity is shown).  (b) Merging behavior in the amplitude equation \eqref{eqn:amplitudeequation} [$\Re A(\xbar,t)$ is shown].}
		\label{fig:merging_jets}
	\end{figure}

The \zf{} behavior in numerical simulations of \eqnref{GHMmodel} is shown in Fig.~\ref{fig:merging_jets}(a).  During the transient period, merging jets are observed, while in the late time a statistically steady state is reached with stable unwavering jets.

We restrict ourselves to the quasilinear (QL) approximation of this system.  To obtain the QL equations, we perform an eddy--mean decomposition, given by decomposing all fields into a zonal mean and a deviation from the zonal mean, then neglect the eddy--eddy nonlinearities within the eddy equation.\cite{srinivasan:2012}  The QL approximation is not expected to be physically and quantitatively correct in detail (though it may be in certain regimes \cite{bouchet:2013}); for example, material conservation of potential vorticity (in the undamped, undriven case) is lost.  However, the QL model is useful because it exhibits the same basic zonal jet features as the full model, namely merging jets and the formation of stable jets.  Therefore, analysis of the QL model can provide a mathematical foundation for understanding and interpreting the physical behavior.

We consider a statistical average of the QL system.  In the presence of steady \zfs{}, a statistical homogeneity assumption is clearly invalid.  Therefore, we allow the turbulence to be inhomogeneous in the direction ($x$) of \zf{} variation.  The averaged equations, referred to as the second-order cumulant expansion (CE2), are \cite{tobias:2011,srinivasan:2012}
	\begin{subequations}
		\label{eqn:CE2}
	\begin{align}
		\partial_t W &+ (U_+ - U_-) \partial_y W - (U_+'' - U_-'') \left( \nablabarsq + \frac{1}{4} \partial_\xbar^2 \right) \partial_y C \notag \\
			& + [2\k + (U_+'' + U_-'')] \partial_\xbar \partial_x \partial_y C \notag \\
			& = F - 2\m W - 2 \n D_h W, \label{eqn:CE2W} \\
		\partial_t U &+ \partial_\xbar \partial_x \partial_y C(0,0,\xbar,t) = -\m U - \n (-1)^h \partial_\xbar^{2h} U, \label{eqn:CE2U}
	\end{align}
	\end{subequations}
where $x$ and $y$ represent two-point separations, $\xbar$ represents the two-point average position (if the turbulence were homogeneous, there would be no $\xbar$ dependence), $W(x,y \mid \xbar,t)$ and $C(x,y \mid \xbar,t)$ are the one-time, two-space-point correlation functions of vorticity and potential, $U(\xbar,t)$ is the zonal flow velocity, $U_\pm = U(\xbar \pm x/2, t)$, $\nablabarsq = \partial_x^2 + \partial_y^2 - 1$, $F(x,y)$ is chosen to be isotropic, homogeneous ring forcing, and $D_h $ is a hyperviscosity operator.  There is a linear relation between $W$ and $C$.\cite{srinivasan:2012}

Given the assumption that the stochastic forcing $\xi$ is white (delta-correlated) noise, the only further assumptions necessary for CE2 to be an exact description of the QL model are statistical homogeneity and ergodicity in the zonal ($y$) direction.  This is because the QL model neglects the nonlinear eddy--eddy term that would give rise to a closure problem.  Alternatively, CE2 can be regarded as a truncated statistical closure of the full model.\cite{farrell:2003,farrell:2007,marston:2008,tobias:2011}

The CE2 equations exhibit important symmetries of translation and reflection, given by $\xbar \to \xbar + \de \xbar$, $(x,\xbar) \to (-x, -\xbar)$, $(y, \xbar) \to (-y, -\xbar)$, and $(x,y) \to (-x,-y)$.


Many studies of CE2 have been performed previously.\cite{farrell:2003,farrell:2007,farrell:2009,marston:2008,tobias:2011,tobias:2013}  Numerical simulations of CE2 also exhibit merging jets.\cite{farrell:2007}

For \eqnref{CE2} there always exists a homogeneous equilibrium: $W(x,y) = (2\m + 2\n D_h)^{-1} F$, $U=0$.  This equilibrium is stable in a certain regime of parameters.  As a control parameter such as $\m$ is varied, this homogeneous state becomes zonostrophically unstable.\cite{srinivasan:2012,farrell:2007}  Physically, zonostrophic instability occurs when dissipation is overcome by the mutually reinforcing processes of eddy tilting by zonal flows and production of Reynolds stress forces by tilted eddies.  The eigenmode consists of perturbations spatially periodic in $\xbar$ with zero real frequency,\cite{srinivasan:2012} so that zonostrophic instability arises as a Type I$_s$ instability \cite{cross:1993} of homogeneous turbulence.  Zonostrophic instability within CE2 may be thought of as a variant of modulational instability calculations of ZF generation.

Just beyond the instability threshold, a bifurcation analysis follows a standard procedure and involves a multiscale perturbation expansion about the threshold.  Let $u$ be the state vector relative to the homogeneous equilibrium and let $\epsilon$ be a normalized control parameter.  The expansion proceeds as $u = \epsilon^{1/2} u_1 + \epsilon u_2 + \cdots$.  At first order one finds $u_1 = A(\xbar,t) r + \text{c.c.}$, where $\text{c.c.}$ denotes complex conjugate and $r \sim e^{i q_c \xbar}$ is the eigenmode that is marginally stable at $\epsilon=0$.  One determines a PDE for the complex amplitude $A$ as a solvability condition at third order in the perturbation expansion.  This amplitude equation is constrained by the translation and reflection symmetries to take a universal form.\cite{cross:1993}  The amplitude equation, sometimes referred to as the real Ginzburg-Landau equation, is
	\begin{equation}
		\partial_t A(\xbar,t) = A + \partial_\xbar^2 A - |A|^2 A,
		\label{eqn:amplitudeequation}
	\end{equation}
where all coefficients have been rescaled to unity.  The derivation of Eq.~\eqref{eqn:amplitudeequation} from Eq.~\eqref{eqn:CE2} will be reported elsewhere.\cite{parker:2013}

The amplitude equation \eqref{eqn:amplitudeequation} is well understood.\cite{cross:1993,cross:2009,hoyle:2006}  First, a steady-state solution exists for any wave number within the continuous band $-1 < k < 1$ (to see this, observe that $A = \a e^{ik\xbar}$ with $|\a|^2 = 1-k^2$ is a solution).  Second, only solutions with $k^2 < 1/3$ are linearly stable.\cite{cross:2009}  This is demonstrated in Fig.~\ref{fig:merging_jets}(b), where an unstable solution that has been slightly perturbed undergoes merging behavior until a stable wave number is reached.  The preceding qualitative behaviors are also exhibited by the CE2 system, as we now show.

We proceed to find the steady-state solutions of \eqnref{CE2}.  In the context of an infinite domain with no boundaries, these solutions are referred to as ideal states.  Let $q$ denote the basic \zf{} wave number of an ideal state.  For a given $q$, we solve the time-independent form of \eqnref{CE2} directly.  This approach is distinct from time integration of \eqnref{CE2} to a steady state.  Our procedure has two advantages for understanding the global structure of the system.  First, we can specify precisely the $q$ of the desired solution.  Second, we can solve directly for all solutions, including unstable ones, rather than find only those which develop from time evolution.

An ideal state is represented as a Fourier-Galerkin series with coefficients to be determined\cite{clever:1974,newell:1990,cross:2009}:
	\begin{subequations}
	\begin{align}
		W(x,y \mid \xbar) &= \sum_{m=-M}^M \sum_{n=-N}^N \sum_{p=-P}^P W_{mnp} e^{imax} e^{inby} e^{ipq\xbar}, \\
		U(\xbar) &= \sum_{p=-P}^P U_p e^{ipq\xbar}.
	\end{align}
	\end{subequations}
While the periodicity in $\xbar$ is desired, the correlation function should decay in $x$ and $y$; periodicity in $x$ and $y$ arsies from using the convenient Fourier basis.  Thus, $a$ and $b$,  unlike $q$, are numerical parameters.  They represent the spectral resolution of the correlation function and should be small enough to obtain an accurate solution.
	
The CE2 symmetries allow us to seek a solution where $U(\xbar) = U(-\xbar)$ and $W(x,y \mid \xbar) = W(-x,-y \mid \xbar) = W(x, -y \mid -\xbar) = W(-x, y \mid -\xbar)$.  These constraints, along with reality conditions, force $U_p$ to be real, $U_p = U_{-p}$, and $W_{mnp} = W_{-m,n,p}^* = W_{m,-n,p}^* = W_{m,n,-p}^*$.

We obtain a system of nonlinear algebraic equations for the coefficients $U_p, W_{mnp}$ by substituting the Galerkin series into \eqnref{CE2} and projecting onto the basis functions.  To demonstrate the projection for \eqnref{CE2W}, let $\p_{mnp} = e^{imax} e^{inby} e^{ipq\xbar}$.  We project \eqnref{CE2W} onto $\p_{rst}$ by operating with 
	\begin{equation}
		\left(\frac{2\pi}{a} \frac{2\pi}{b} \frac{2\pi}{q}\right)^{-1}  \int_{-\pi/a}^{\pi/a} dx   \int_{-\pi/b}^{\pi/b} dy 	\int_{-\pi/q}^{\pi/q} d\xbar\,  \p_{rst}^*.
	\end{equation}
For instance, the term $(U_+ - U_-) \partial_y W$ projects to $I_{rstp'mnp} U_{p'} W_{mnp}$, where repeated indices are summed over, $I_{rstp'mnp} = inb \de_{n,s} \de_{p'+p-t,0} (\s_+ - \s_-)$, $\s_\pm = \sinc( \a_\pm \pi/a )$, and $\a_\pm = ma - ra \pm p'q/2$.  The other terms of \eqnref{CE2W}, as well as \eqnref{CE2U}, are handled similarly.
	
\begin{figure}
		\centering
		\includegraphics{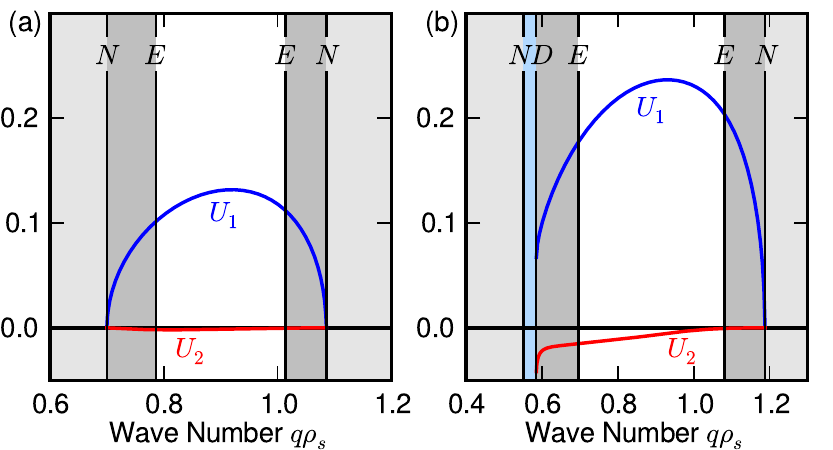}
		\caption{Zonal flow amplitude $U_1$, $U_2$ as a function of ideal state wave number $q$ at (a) $\m=0.21$ ($R_\b =1.48$) and (b) $\m=0.19$ ($R_\b=1.51$).  In the unshaded region, ideal states are stable.  The vertical lines correspond to various instabilities which separate the regions (see Fig. \ref{fig:stability_balloon}).}
		\label{fig:zf_amplitude}
	\end{figure}

\begin{figure}
		\centering
		\includegraphics{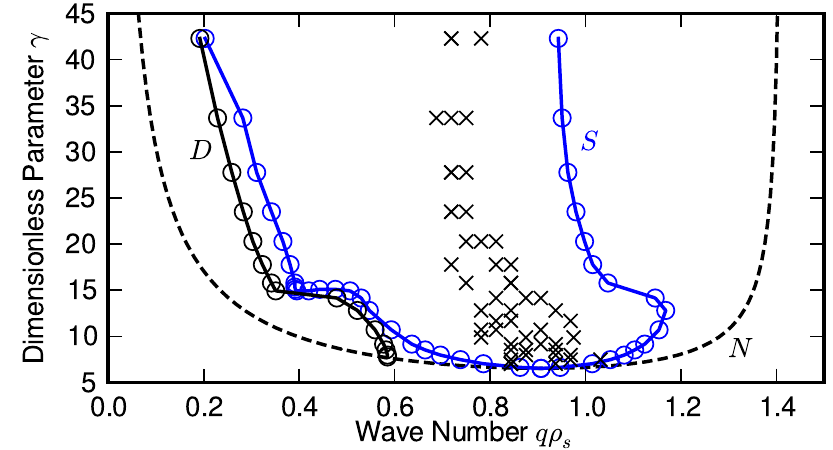}
		\caption{Stability diagram for the CE2 equations.  Above the neutral curve (N), the homogeneous turbulent state is zonostrophically unstable.  Ideal states are stable within the marginal stability curve $S$.  The stability curve is consistent with the dominant \zf{} wavenumber from independent QL simulations (crosses).  The stationary ideal states vanish to the left of $D$.  Here, $a=0.06$, $b=0.08$, $M=20$, $N=33$, $P=5$, and other parameters are given in the text.  $\gamma $ is varied by changing $\m$ while holding other parameters fixed.}
		\label{fig:stability_balloon}
	\end{figure}

The system of nonlinear algebraic equations is solved with a Newton's method.\cite{kelley:2003}  Figure \ref{fig:zf_amplitude} shows the \zf{} amplitude coefficients $U_p$ as functions of $q$ at $\m = 0.21$ and $\m = 0.19$.  Near the instability threshold, ideal states exist at all $q$ for which the homogeneous equilibrium is zonostrophically unstable [between the two lines labeled $N$ in Fig.~\ref{fig:zf_amplitude}(a)].  Farther from threshold, there is a region of $q$ where the ideal-state solution seems to disappear [between the lines $N$ and $D$ in Fig. \ref{fig:zf_amplitude}(b); see also Fig. \ref{fig:stability_balloon}].  The values of the other parameters used are $\k=1$, $\n=10^{-3}$, and $h=4$.  The forcing $F(\v{k}) = 2\pi \ve k_f / \delta k$ for $k_f - \delta k < |\v{k}| < k_f + \delta k$, and is zero otherwise.  We take $k_f=1$, $\delta k = 1/8$, and $\ve$, which acts like a total energy input rate, to be equal to 1.

To investigate stability of the ideal states, we consider perturbations $\de W(x,y \mid \xbar, t)$ and $\de U(\xbar, t)$ about an equilibrium $W, U$ and linearize Eq.~\eqref{eqn:CE2}.  Since the underlying equilibrium is periodic in $\xbar$, the perturbations can be expanded as a Bloch state\cite{cross:2009, clever:1974}:
	\begin{subequations}
	\begin{align}
		\de W(x,y\mid \xbar , t) &= e^{\s t} e^{iQ\xbar} \sum_{mnp} \de W_{mnp} e^{imax} e^{inby} e^{ipq\xbar}, \\
		\de U(\xbar, t) &= e^{\s t} e^{iQ\xbar} \sum_p \de U_p e^{ipq\xbar},
	\end{align}
	\end{subequations}
where $Q$ is the Bloch wave number and can be taken to lie within the first Brillouin zone $-q/2 < Q \le q/2$.  We do not use a $Q_x$ or $Q_y$ because as previously mentioned the periodicity in $x$ and $y$ is artificial.  The perturbation equations are projected onto the basis functions in the same way as in the ideal state calculation.  This projection results in a linear system at each $Q$ for the coefficients $\de W_{mnp}$ and $\de U_p$; this determines an eigenvalue problem for $\s$.  The equilibrium is unstable if for any $Q$ there are any eigenvalues with $\Re \s > 0$.

The stability diagram is shown in Fig.~\ref{fig:stability_balloon}.  As the control parameter we adopt $\gamma = \ve^{1/4} \k^{1/2} \m^{-5/4}$, an important dimensionless parameter controlling the \zf{} dynamics.\cite{galperin:2010,danilov:2004}  To vary $\gamma$, we change $\m$ and hold other parameters fixed at their previous values.  The stable ideal states exist inside of the marginal stability curve marked $S$.  Near the threshold, marginal stability is governed by the Eckhaus instability, a long-wavelength universal instability.\cite{cross:1993}  Farther from threshold, the instability transitions into new, nonuniversal instabilities; details will be reported elsewhere.  The \zfs{} are spontaneously generated for $\gamma > 6.53$.  For $\gamma > 6.53$, the stability curve is consistent with the dominant \zf{} wavenumber observed in QL simulations.

Numerical simulations typically are done within a finite domain.  When periodic boundary conditions are used, our infinite-domain results are modified merely by the discretization of wave numbers.  This affects not only the possible equilibria, but also any perturbations and hence the stability boundaries too.  

For a time-evolving system, the exact $q$ that is ultimately chosen within the stability balloon results from a dynamical process and is not addressed in a systematic way by the present study.

While the CE2 equations exhibit spontaneously generated zonal flows, they neglect many physical effects.  An important piece of physics missing from the CE2 equations is the nonlinear eddy self-interaction, which clearly cannot be ignored in general.  At least one particular instance of the qualitative failure of CE2 has been noted.\cite{tobias:2013}

Yet, the basic mathematical structure of the theory presented here arises only from symmetry arguments and general properties of the zonostrophic instability.  If one were to include the important physics neglected in CE2, those general symmetries and properties should remain intact.  Therefore, we expect our qualitative conclusions to likewise remain valid.

In summary, by analyzing a second-order statistical model of an ensemble of interacting zonal flows and turbulence, we have shown that zonal flows constitute pattern formation amid a turbulent bath.  This continues previous work\cite{srinivasan:2012} to provide a firm analytic understanding of zonal flow generation and equilibrium within CE2.  We calculated the stability diagram of steady zonal jets and explained the merging of jets as a means of attaining a stable wave number.  In general, the use of statistically averaged equations and the pattern formation methodology provide a path forward for further systematic investigations of zonal flows and their interactions with turbulence.  Further work should be done to understand how this framework can shed light on practical problems involving realistic plasmas in the near-collisionless regime.

We acknowledge useful discussions with Brian Farrell, Henry Greenside, Petros Ioannou, and Brad Marston.  This material is based upon work supported by an NSF Graduate Research Fellowship and a US DOE Fusion Energy Sciences Fellowship.  This work was also supported by US DOE Contract DE-AC02-09CH11466.


\begin{thebibliography}{36}%
\makeatletter
\providecommand \@ifxundefined [1]{%
 \@ifx{#1\undefined}
}%
\providecommand \@ifnum [1]{%
 \ifnum #1\expandafter \@firstoftwo
 \else \expandafter \@secondoftwo
 \fi
}%
\providecommand \@ifx [1]{%
 \ifx #1\expandafter \@firstoftwo
 \else \expandafter \@secondoftwo
 \fi
}%
\providecommand \natexlab [1]{#1}%
\providecommand \enquote  [1]{``#1''}%
\providecommand \bibnamefont  [1]{#1}%
\providecommand \bibfnamefont [1]{#1}%
\providecommand \citenamefont [1]{#1}%
\providecommand \href@noop [0]{\@secondoftwo}%
\providecommand \href [0]{\begingroup \@sanitize@url \@href}%
\providecommand \@href[1]{\@@startlink{#1}\@@href}%
\providecommand \@@href[1]{\endgroup#1\@@endlink}%
\providecommand \@sanitize@url [0]{\catcode `\\12\catcode `\$12\catcode
  `\&12\catcode `\#12\catcode `\^12\catcode `\_12\catcode `\%12\relax}%
\providecommand \@@startlink[1]{}%
\providecommand \@@endlink[0]{}%
\providecommand \url  [0]{\begingroup\@sanitize@url \@url }%
\providecommand \@url [1]{\endgroup\@href {#1}{\urlprefix }}%
\providecommand \urlprefix  [0]{URL }%
\providecommand \Eprint [0]{\href }%
\providecommand \doibase [0]{http://dx.doi.org/}%
\providecommand \selectlanguage [0]{\@gobble}%
\providecommand \bibinfo  [0]{\@secondoftwo}%
\providecommand \bibfield  [0]{\@secondoftwo}%
\providecommand \translation [1]{[#1]}%
\providecommand \BibitemOpen [0]{}%
\providecommand \bibitemStop [0]{}%
\providecommand \bibitemNoStop [0]{.\EOS\space}%
\providecommand \EOS [0]{\spacefactor3000\relax}%
\providecommand \BibitemShut  [1]{\csname bibitem#1\endcsname}%
\let\auto@bib@innerbib\@empty
\bibitem [{\citenamefont {Vasavada}\ and\ \citenamefont
  {Showman}(2005)}]{vasavada:2005}%
  \BibitemOpen
  \bibfield  {author} {\bibinfo {author} {\bibfnamefont {A.~R.}\ \bibnamefont
  {Vasavada}}\ and\ \bibinfo {author} {\bibfnamefont {A.~P.}\ \bibnamefont
  {Showman}},\ }\href {http://stacks.iop.org/0034-4885/68/i=8/a=R06} {\bibfield
   {journal} {\bibinfo  {journal} {Rep. Prog. Phys.}\ }\textbf {\bibinfo
  {volume} {68}},\ \bibinfo {pages} {1935} (\bibinfo {year}
  {2005})}\BibitemShut {NoStop}%
\bibitem [{\citenamefont {Fujisawa}(2009)}]{fujisawa:2009}%
  \BibitemOpen
  \bibfield  {author} {\bibinfo {author} {\bibfnamefont {A.}~\bibnamefont
  {Fujisawa}},\ }\href {http://stacks.iop.org/0029-5515/49/i=1/a=013001}
  {\bibfield  {journal} {\bibinfo  {journal} {Nucl. Fusion}\ }\textbf {\bibinfo
  {volume} {49}},\ \bibinfo {pages} {013001} (\bibinfo {year}
  {2009})}\BibitemShut {NoStop}%
\bibitem [{\citenamefont {Johansen}\ \emph {et~al.}(2009)\citenamefont
  {Johansen}, \citenamefont {Youdin},\ and\ \citenamefont
  {Klahr}}]{johansen:2009}%
  \BibitemOpen
  \bibfield  {author} {\bibinfo {author} {\bibfnamefont {A.}~\bibnamefont
  {Johansen}}, \bibinfo {author} {\bibfnamefont {A.}~\bibnamefont {Youdin}}, \
  and\ \bibinfo {author} {\bibfnamefont {H.}~\bibnamefont {Klahr}},\ }\href
  {http://stacks.iop.org/0004-637X/697/i=2/a=1269} {\bibfield  {journal}
  {\bibinfo  {journal} {Astrophys. J.}\ }\textbf {\bibinfo {volume} {697}},\
  \bibinfo {pages} {1269} (\bibinfo {year} {2009})}\BibitemShut {NoStop}%
\bibitem [{\citenamefont {Lin}\ \emph {et~al.}(1998)\citenamefont {Lin},
  \citenamefont {Hahm}, \citenamefont {Lee}, \citenamefont {Tang},\ and\
  \citenamefont {White}}]{lin:1998}%
  \BibitemOpen
  \bibfield  {author} {\bibinfo {author} {\bibfnamefont {Z.}~\bibnamefont
  {Lin}}, \bibinfo {author} {\bibfnamefont {T.~S.}\ \bibnamefont {Hahm}},
  \bibinfo {author} {\bibfnamefont {W.~W.}\ \bibnamefont {Lee}}, \bibinfo
  {author} {\bibfnamefont {W.~M.}\ \bibnamefont {Tang}}, \ and\ \bibinfo
  {author} {\bibfnamefont {R.~B.}\ \bibnamefont {White}},\ }\href {\doibase
  10.1126/science.281.5384.1835} {\bibfield  {journal} {\bibinfo  {journal}
  {Science}\ }\textbf {\bibinfo {volume} {281}},\ \bibinfo {pages} {1835}
  (\bibinfo {year} {1998})}\BibitemShut {NoStop}%
\bibitem [{\citenamefont {Diamond}\ \emph {et~al.}(2005)\citenamefont
  {Diamond}, \citenamefont {Itoh}, \citenamefont {Itoh},\ and\ \citenamefont
  {Hahm}}]{diamond:2005}%
  \BibitemOpen
  \bibfield  {author} {\bibinfo {author} {\bibfnamefont {P.~H.}\ \bibnamefont
  {Diamond}}, \bibinfo {author} {\bibfnamefont {S.-I.}\ \bibnamefont {Itoh}},
  \bibinfo {author} {\bibfnamefont {K.}~\bibnamefont {Itoh}}, \ and\ \bibinfo
  {author} {\bibfnamefont {T.~S.}\ \bibnamefont {Hahm}},\ }\href
  {http://stacks.iop.org/0741-3335/47/i=5/a=R01} {\bibfield  {journal}
  {\bibinfo  {journal} {Plasma Physics and Controlled Fusion}\ }\textbf
  {\bibinfo {volume} {47}},\ \bibinfo {pages} {R35} (\bibinfo {year}
  {2005})}\BibitemShut {NoStop}%
\bibitem [{\citenamefont {Rogers}\ \emph {et~al.}(2000)\citenamefont {Rogers},
  \citenamefont {Dorland},\ and\ \citenamefont
  {Kotschenreuther}}]{rogers:2000}%
  \BibitemOpen
  \bibfield  {author} {\bibinfo {author} {\bibfnamefont {B.~N.}\ \bibnamefont
  {Rogers}}, \bibinfo {author} {\bibfnamefont {W.}~\bibnamefont {Dorland}}, \
  and\ \bibinfo {author} {\bibfnamefont {M.}~\bibnamefont {Kotschenreuther}},\
  }\href {\doibase 10.1103/PhysRevLett.85.5336} {\bibfield  {journal} {\bibinfo
   {journal} {Phys. Rev. Lett.}\ }\textbf {\bibinfo {volume} {85}},\ \bibinfo
  {pages} {5336} (\bibinfo {year} {2000})}\BibitemShut {NoStop}%
\bibitem [{\citenamefont {Connaughton}\ \emph {et~al.}(2011)\citenamefont
  {Connaughton}, \citenamefont {Nazarenko},\ and\ \citenamefont
  {Quinn}}]{connaughton:2011}%
  \BibitemOpen
  \bibfield  {author} {\bibinfo {author} {\bibfnamefont {C.}~\bibnamefont
  {Connaughton}}, \bibinfo {author} {\bibfnamefont {S.}~\bibnamefont
  {Nazarenko}}, \ and\ \bibinfo {author} {\bibfnamefont {B.}~\bibnamefont
  {Quinn}},\ }\href {http://stacks.iop.org/0295-5075/96/i=2/a=25001} {\bibfield
   {journal} {\bibinfo  {journal} {EPL}\ }\textbf {\bibinfo {volume} {96}},\
  \bibinfo {pages} {25001} (\bibinfo {year} {2011})}\BibitemShut {NoStop}%
\bibitem [{\citenamefont {Rhines}(1975)}]{rhines:1975}%
  \BibitemOpen
  \bibfield  {author} {\bibinfo {author} {\bibfnamefont {P.~B.}\ \bibnamefont
  {Rhines}},\ }\href@noop {} {\bibfield  {journal} {\bibinfo  {journal} {J.
  Fluid Mech.}\ }\textbf {\bibinfo {volume} {69}},\ \bibinfo {pages} {417}
  (\bibinfo {year} {1975})}\BibitemShut {NoStop}%
\bibitem [{\citenamefont {Vallis}\ and\ \citenamefont
  {Maltrud}(1993)}]{vallis:1993}%
  \BibitemOpen
  \bibfield  {author} {\bibinfo {author} {\bibfnamefont {G.~K.}\ \bibnamefont
  {Vallis}}\ and\ \bibinfo {author} {\bibfnamefont {M.~E.}\ \bibnamefont
  {Maltrud}},\ }\href {\doibase
  10.1175/1520-0485(1993)023<1346:GOMFAJ>2.0.CO;2} {\bibfield  {journal}
  {\bibinfo  {journal} {J. Phys. Oceanogr.}\ }\textbf {\bibinfo {volume}
  {23}},\ \bibinfo {pages} {1346} (\bibinfo {year} {1993})}\BibitemShut
  {NoStop}%
\bibitem [{\citenamefont {Sukoriansky}\ \emph {et~al.}(2007)\citenamefont
  {Sukoriansky}, \citenamefont {Dikovskaya},\ and\ \citenamefont
  {Galperin}}]{sukoriansky:2007}%
  \BibitemOpen
  \bibfield  {author} {\bibinfo {author} {\bibfnamefont {S.}~\bibnamefont
  {Sukoriansky}}, \bibinfo {author} {\bibfnamefont {N.}~\bibnamefont
  {Dikovskaya}}, \ and\ \bibinfo {author} {\bibfnamefont {B.}~\bibnamefont
  {Galperin}},\ }\href {\doibase 10.1175/JAS4013.1} {\bibfield  {journal}
  {\bibinfo  {journal} {J. Atmos. Sci.}\ }\textbf {\bibinfo {volume} {64}},\
  \bibinfo {pages} {3312} (\bibinfo {year} {2007})}\BibitemShut {NoStop}%
\bibitem [{\citenamefont {Dritschel}\ and\ \citenamefont
  {McIntyre}(2008)}]{dritschel:2008}%
  \BibitemOpen
  \bibfield  {author} {\bibinfo {author} {\bibfnamefont {D.~G.}\ \bibnamefont
  {Dritschel}}\ and\ \bibinfo {author} {\bibfnamefont {M.~E.}\ \bibnamefont
  {McIntyre}},\ }\href {\doibase 10.1175/2007JAS2227.1} {\bibfield  {journal}
  {\bibinfo  {journal} {J. Atmos. Sci.}\ }\textbf {\bibinfo {volume} {65}},\
  \bibinfo {pages} {855} (\bibinfo {year} {2008})}\BibitemShut {NoStop}%
\bibitem [{\citenamefont {Scott}\ and\ \citenamefont
  {Dritschel}(2012)}]{scott:2012}%
  \BibitemOpen
  \bibfield  {author} {\bibinfo {author} {\bibfnamefont {R.~K.}\ \bibnamefont
  {Scott}}\ and\ \bibinfo {author} {\bibfnamefont {D.~G.}\ \bibnamefont
  {Dritschel}},\ }\href {\doibase 10.1017/jfm.2012.410} {\bibfield  {journal}
  {\bibinfo  {journal} {Journal of Fluid Mechanics}\ }\textbf {\bibinfo
  {volume} {711}},\ \bibinfo {pages} {576} (\bibinfo {year}
  {2012})}\BibitemShut {NoStop}%
\bibitem [{\citenamefont {Huang}\ and\ \citenamefont
  {Robinson}(1998)}]{huang:1998}%
  \BibitemOpen
  \bibfield  {author} {\bibinfo {author} {\bibfnamefont {H.-P.}\ \bibnamefont
  {Huang}}\ and\ \bibinfo {author} {\bibfnamefont {W.~A.}\ \bibnamefont
  {Robinson}},\ }\href {\doibase
  10.1175/1520-0469(1998)055<0611:TDTAPZ>2.0.CO;2} {\bibfield  {journal}
  {\bibinfo  {journal} {J. Atmos. Sci.}\ }\textbf {\bibinfo {volume} {55}},\
  \bibinfo {pages} {611} (\bibinfo {year} {1998})}\BibitemShut {NoStop}%
\bibitem [{\citenamefont {Scott}\ and\ \citenamefont
  {Polvani}(2007)}]{scott:2007}%
  \BibitemOpen
  \bibfield  {author} {\bibinfo {author} {\bibfnamefont {R.~K.}\ \bibnamefont
  {Scott}}\ and\ \bibinfo {author} {\bibfnamefont {L.~M.}\ \bibnamefont
  {Polvani}},\ }\href {\doibase 10.1175/JAS4003.1} {\bibfield  {journal}
  {\bibinfo  {journal} {J. Atmos. Sci.}\ }\textbf {\bibinfo {volume} {64}},\
  \bibinfo {pages} {3158} (\bibinfo {year} {2007})}\BibitemShut {NoStop}%
\bibitem [{\citenamefont {Smolyakov}\ \emph {et~al.}(2000)\citenamefont
  {Smolyakov}, \citenamefont {Diamond},\ and\ \citenamefont
  {Malkov}}]{smolyakov:2000}%
  \BibitemOpen
  \bibfield  {author} {\bibinfo {author} {\bibfnamefont {A.~I.}\ \bibnamefont
  {Smolyakov}}, \bibinfo {author} {\bibfnamefont {P.~H.}\ \bibnamefont
  {Diamond}}, \ and\ \bibinfo {author} {\bibfnamefont {M.}~\bibnamefont
  {Malkov}},\ }\href {\doibase 10.1103/PhysRevLett.84.491} {\bibfield
  {journal} {\bibinfo  {journal} {Phys. Rev. Lett.}\ }\textbf {\bibinfo
  {volume} {84}},\ \bibinfo {pages} {491} (\bibinfo {year} {2000})}\BibitemShut
  {NoStop}%
\bibitem [{\citenamefont {Krommes}\ and\ \citenamefont
  {Kim}(2000)}]{krommes:2000}%
  \BibitemOpen
  \bibfield  {author} {\bibinfo {author} {\bibfnamefont {J.~A.}\ \bibnamefont
  {Krommes}}\ and\ \bibinfo {author} {\bibfnamefont {C.-B.}\ \bibnamefont
  {Kim}},\ }\href {\doibase 10.1103/PhysRevE.62.8508} {\bibfield  {journal}
  {\bibinfo  {journal} {Phys. Rev. E}\ }\textbf {\bibinfo {volume} {62}},\
  \bibinfo {pages} {8508} (\bibinfo {year} {2000})}\BibitemShut {NoStop}%
\bibitem [{\citenamefont {Srinivasan}\ and\ \citenamefont
  {Young}(2012)}]{srinivasan:2012}%
  \BibitemOpen
  \bibfield  {author} {\bibinfo {author} {\bibfnamefont {K.}~\bibnamefont
  {Srinivasan}}\ and\ \bibinfo {author} {\bibfnamefont {W.~R.}\ \bibnamefont
  {Young}},\ }\href {\doibase 10.1175/JAS-D-11-0200.1} {\bibfield  {journal}
  {\bibinfo  {journal} {J. Atmos. Sci.}\ }\textbf {\bibinfo {volume} {69}},\
  \bibinfo {pages} {1633} (\bibinfo {year} {2012})}\BibitemShut {NoStop}%
\bibitem [{\citenamefont {Farrell}\ and\ \citenamefont
  {Ioannou}(2003)}]{farrell:2003}%
  \BibitemOpen
  \bibfield  {author} {\bibinfo {author} {\bibfnamefont {B.~F.}\ \bibnamefont
  {Farrell}}\ and\ \bibinfo {author} {\bibfnamefont {P.~J.}\ \bibnamefont
  {Ioannou}},\ }\href
  {http://dx.doi.org/10.1175/1520-0469(2003)060<2101:SSOTJ>2.0.CO;2} {\bibfield
   {journal} {\bibinfo  {journal} {J. Atmos. Sci.}\ }\textbf {\bibinfo {volume}
  {60}},\ \bibinfo {pages} {2101} (\bibinfo {year} {2003})}\BibitemShut
  {NoStop}%
\bibitem [{\citenamefont {Farrell}\ and\ \citenamefont
  {Ioannou}(2007)}]{farrell:2007}%
  \BibitemOpen
  \bibfield  {author} {\bibinfo {author} {\bibfnamefont {B.~F.}\ \bibnamefont
  {Farrell}}\ and\ \bibinfo {author} {\bibfnamefont {P.~J.}\ \bibnamefont
  {Ioannou}},\ }\href {\doibase 10.1175/JAS4016.1} {\bibfield  {journal}
  {\bibinfo  {journal} {J. Atmos. Sci.}\ }\textbf {\bibinfo {volume} {64}},\
  \bibinfo {pages} {3652} (\bibinfo {year} {2007})}\BibitemShut {NoStop}%
\bibitem [{\citenamefont {Bakas}\ and\ \citenamefont
  {Ioannou}(2013)}]{bakas:2013}%
  \BibitemOpen
  \bibfield  {author} {\bibinfo {author} {\bibfnamefont {N.~A.}\ \bibnamefont
  {Bakas}}\ and\ \bibinfo {author} {\bibfnamefont {P.~J.}\ \bibnamefont
  {Ioannou}},\ }\href {\doibase 10.1103/PhysRevLett.110.224501} {\bibfield
  {journal} {\bibinfo  {journal} {Phys. Rev. Lett.}\ }\textbf {\bibinfo
  {volume} {110}},\ \bibinfo {pages} {224501} (\bibinfo {year}
  {2013})}\BibitemShut {NoStop}%
\bibitem{constantinou:2012}
N.~C.~Constantinou, P.~J.~Ioannou, and B.~F.~Farrell, \emph{Emergence and equilibration of jets in beta-plane turbulence: applications of Stochastic Structural Stability Theory}, submitted to J.~Atmos.~Sci.
\bibitem [{\citenamefont {Marston}\ \emph {et~al.}(2008)\citenamefont
  {Marston}, \citenamefont {Conover},\ and\ \citenamefont
  {Schneider}}]{marston:2008}%
  \BibitemOpen
  \bibfield  {author} {\bibinfo {author} {\bibfnamefont {J.~B.}\ \bibnamefont
  {Marston}}, \bibinfo {author} {\bibfnamefont {E.}~\bibnamefont {Conover}}, \
  and\ \bibinfo {author} {\bibfnamefont {T.}~\bibnamefont {Schneider}},\ }\href
  {\doibase 10.1175/2007JAS2510.1} {\bibfield  {journal} {\bibinfo  {journal}
  {J. Atmos. Sci.}\ }\textbf {\bibinfo {volume} {65}},\ \bibinfo {pages} {1955}
  (\bibinfo {year} {2008})}\BibitemShut {NoStop}%
\bibitem [{\citenamefont {Tobias}\ \emph {et~al.}(2011)\citenamefont {Tobias},
  \citenamefont {Dagon},\ and\ \citenamefont {Marston}}]{tobias:2011}%
  \BibitemOpen
  \bibfield  {author} {\bibinfo {author} {\bibfnamefont {S.~M.}\ \bibnamefont
  {Tobias}}, \bibinfo {author} {\bibfnamefont {K.}~\bibnamefont {Dagon}}, \
  and\ \bibinfo {author} {\bibfnamefont {J.~B.}\ \bibnamefont {Marston}},\
  }\href {http://stacks.iop.org/0004-637X/727/i=2/a=127} {\bibfield  {journal}
  {\bibinfo  {journal} {Astrophys. J.}\ }\textbf {\bibinfo {volume} {727}},\
  \bibinfo {pages} {127} (\bibinfo {year} {2011})}\BibitemShut {NoStop}%
\bibitem [{\citenamefont {Tobias}\ and\ \citenamefont
  {Marston}(2013)}]{tobias:2013}%
  \BibitemOpen
  \bibfield  {author} {\bibinfo {author} {\bibfnamefont {S.~M.}\ \bibnamefont
  {Tobias}}\ and\ \bibinfo {author} {\bibfnamefont {J.~B.}\ \bibnamefont
  {Marston}},\ }\href {\doibase 10.1103/PhysRevLett.110.104502} {\bibfield
  {journal} {\bibinfo  {journal} {Phys. Rev. Lett.}\ }\textbf {\bibinfo
  {volume} {110}},\ \bibinfo {pages} {104502} (\bibinfo {year}
  {2013})}\BibitemShut {NoStop}%
\bibitem [{\citenamefont {Krommes}(2002)}]{krommes:2002}%
  \BibitemOpen
  \bibfield  {author} {\bibinfo {author} {\bibfnamefont {J.~A.}\ \bibnamefont
  {Krommes}},\ }\href
  {http://www.sciencedirect.com/science/article/pii/S0370157301000667}
  {\bibfield  {journal} {\bibinfo  {journal} {Phys. Rep.}\ }\textbf {\bibinfo
  {volume} {360}},\ \bibinfo {pages} {1} (\bibinfo {year} {2002})}\BibitemShut
  {NoStop}%
\bibitem [{\citenamefont {Cross}\ and\ \citenamefont
  {Hohenberg}(1993)}]{cross:1993}%
  \BibitemOpen
  \bibfield  {author} {\bibinfo {author} {\bibfnamefont {M.~C.}\ \bibnamefont
  {Cross}}\ and\ \bibinfo {author} {\bibfnamefont {P.~C.}\ \bibnamefont
  {Hohenberg}},\ }\href {\doibase 10.1103/RevModPhys.65.851} {\bibfield
  {journal} {\bibinfo  {journal} {Rev. Mod. Phys.}\ }\textbf {\bibinfo {volume}
  {65}},\ \bibinfo {pages} {851} (\bibinfo {year} {1993})}\BibitemShut
  {NoStop}%
\bibitem [{\citenamefont {Cross}\ and\ \citenamefont
  {Greenside}(2009)}]{cross:2009}%
  \BibitemOpen
  \bibfield  {author} {\bibinfo {author} {\bibfnamefont {M.}~\bibnamefont
  {Cross}}\ and\ \bibinfo {author} {\bibfnamefont {H.}~\bibnamefont
  {Greenside}},\ }\href
  {http://www.cambridge.org/catalogue/catalogue.asp?isbn=9780521770507} {\emph
  {\bibinfo {title} {Pattern Formation and Dynamics in Nonequilibrium
  Systems}}}\ (\bibinfo  {publisher} {Cambridge University Press, Cambridge},\ \bibinfo
  {year} {2009})\BibitemShut {NoStop}%
\bibitem [{\citenamefont {Hoyle}(2006)}]{hoyle:2006}%
  \BibitemOpen
  \bibfield  {author} {\bibinfo {author} {\bibfnamefont {R.}~\bibnamefont
  {Hoyle}},\ }\href {http://books.google.com/books?id=4ZR44NBVCkcC} {\emph
  {\bibinfo {title} {Pattern Formation: An Introduction to Methods}}}\
  (\bibinfo  {publisher} {Cambridge University Press, Cambridge},\ \bibinfo {year}
  {2006})\BibitemShut {NoStop}%
\bibitem [{\citenamefont {Clever}\ and\ \citenamefont
  {Busse}(1974)}]{clever:1974}%
  \BibitemOpen
  \bibfield  {author} {\bibinfo {author} {\bibfnamefont {R.~M.}\ \bibnamefont
  {Clever}}\ and\ \bibinfo {author} {\bibfnamefont {F.~H.}\ \bibnamefont
  {Busse}},\ }\href {\doibase 10.1017/S0022112074001571} {\bibfield  {journal}
  {\bibinfo  {journal} {J. Fluid Mech.}\ }\textbf {\bibinfo {volume} {65}},\
  \bibinfo {pages} {625} (\bibinfo {year} {1974})}\BibitemShut {NoStop}%
\bibitem [{\citenamefont {Newell}\ \emph {et~al.}(1990)\citenamefont {Newell},
  \citenamefont {Passot},\ and\ \citenamefont {Souli}}]{newell:1990}%
  \BibitemOpen
  \bibfield  {author} {\bibinfo {author} {\bibfnamefont {A.~C.}\ \bibnamefont
  {Newell}}, \bibinfo {author} {\bibfnamefont {T.}~\bibnamefont {Passot}}, \
  and\ \bibinfo {author} {\bibfnamefont {M.}~\bibnamefont {Souli}},\ }\href
  {\doibase 10.1017/S0022112090003238} {\bibfield  {journal} {\bibinfo
  {journal} {J. Fluid Mech.}\ }\textbf {\bibinfo {volume} {220}},\ \bibinfo
  {pages} {187} (\bibinfo {year} {1990})}\BibitemShut {NoStop}%
\bibitem{bouchet:2013}
F.~Bouchet, C.~Nardini, and T.~Tangarife, J.~Stat.~Phys. (2013).
\bibitem [{\citenamefont {Farrell}\ and\ \citenamefont
  {Ioannou}(2009)}]{farrell:2009}%
  \BibitemOpen
  \bibfield  {author} {\bibinfo {author} {\bibfnamefont {B.~F.}\ \bibnamefont
  {Farrell}}\ and\ \bibinfo {author} {\bibfnamefont {P.~J.}\ \bibnamefont
  {Ioannou}},\ }\href {\doibase 10.1175/2009JAS2989.1} {\bibfield  {journal}
  {\bibinfo  {journal} {J. Atmos. Sci.}\ }\textbf {\bibinfo {volume} {66}},\
  \bibinfo {pages} {2444} (\bibinfo {year} {2009})}\BibitemShut {NoStop}%
\bibitem{parker:2013}
J.~B.~Parker and J.~A.~Krommes, \emph{Generation of zonal flows through symmetry breaking of statistical homogeneity}, submitted to New.~J.~Phys.
\bibitem [{\citenamefont {Kelley}(2003)}]{kelley:2003}%
  \BibitemOpen
  \bibfield  {author} {\bibinfo {author} {\bibfnamefont {C.~T.}\ \bibnamefont
  {Kelley}},\ }\href@noop {} {\emph {\bibinfo {title} {Solving Nonlinear
  Equations with Newton's Method}}}\ (\bibinfo  {publisher} {Society for
  Industrial and Applied Mathematics, Philadelphia},\ \bibinfo {year} {2003})\BibitemShut
  {NoStop}%
\bibitem [{\citenamefont {Galperin}\ \emph {et~al.}(2010)\citenamefont
  {Galperin}, \citenamefont {Sukoriansky},\ and\ \citenamefont
  {Dikovskaya}}]{galperin:2010}%
  \BibitemOpen
  \bibfield  {author} {\bibinfo {author} {\bibfnamefont {B.}~\bibnamefont
  {Galperin}}, \bibinfo {author} {\bibfnamefont {S.}~\bibnamefont
  {Sukoriansky}}, \ and\ \bibinfo {author} {\bibfnamefont {N.}~\bibnamefont
  {Dikovskaya}},\ }\href {\doibase 10.1007/s10236-010-0278-2} {\bibfield
  {journal} {\bibinfo  {journal} {Ocean Dynamics}\ }\textbf {\bibinfo {volume}
  {60}},\ \bibinfo {pages} {427} (\bibinfo {year} {2010})}\BibitemShut
  {NoStop}%
\bibitem [{\citenamefont {Danilov}\ and\ \citenamefont
  {Gurarie}(2004)}]{danilov:2004}%
  \BibitemOpen
  \bibfield  {author} {\bibinfo {author} {\bibfnamefont {S.}~\bibnamefont
  {Danilov}}\ and\ \bibinfo {author} {\bibfnamefont {D.}~\bibnamefont
  {Gurarie}},\ }\href {\doibase 10.1063/1.1752928} {\bibfield  {journal}
  {\bibinfo  {journal} {Physics of Fluids}\ }\textbf {\bibinfo {volume} {16}},\
  \bibinfo {pages} {2592} (\bibinfo {year} {2004})}\BibitemShut {NoStop}%
\end{thebibliography}

%

\end{document}